\documentstyle[12pt,aasms]{article}
\tightenlines
\begin{document}

\title{Quantitative Analysis of Voids in Percolating Structures in Two-Dimensional N-Body Simulations}
\author{Patrick M. Harrington, Adrian L. Melott, and Sergei F. Shandarin}
\affil{Department of Physics and Astronomy, University of Kansas, Lawrence, KS 66045}

\begin{abstract}

	We present in this paper a quantitative method for defining void size
in large-scale structure based on percolation threshold density.  Beginning 
with two-dimensional gravitational clustering simulations smoothed to the
threshold of nonlinearity, we perform percolation analysis to determine the
large scale structure.  The resulting objective definition 
of voids has a natural
scaling property, is topologically interesting, and can be applied immediately
to redshift surveys.

\end{abstract}

\section{Introduction}

	There has been increasing interest in large voids in the galaxy 
distribution.  However, in large-scale theory and practice, there currently 
exist many different 
definitions of a void (for discussion see Sahni, Sathyaprakash \& Shandarin, 
1993).  Our purpose here is to present an objective and quantitative method 
for defining voids.  We use an algorithm developed by Kauffmann 
(see Kauffmann \& Melott 1992, hereafter KM) to locate voids.  We use a 
percolation technique 
suggested by Shandarin (1983) in the form developed by Klypin \& Shandarin
(1993, hereafter KS) to determine the threshold density.  We perform the
analysis on two-dimensional N-body generated density distributions used in 
the work of 
Beacom et al (1991, hereafter BDMPS).

	The choice of the percolation threshold is motivated by the fact
that it marks a change of topology.  In the absence of percolation,
the regions above the density threshold are isolated.  On the contrary, in
two-dimensional (hereafter 2-D) space, if the fraction above the threshold 
percolates, the fraction
below the threshold forms isolated voids.

	We perform this work in two dimensions because coding 
and de-bugging are
obviously simpler and visualization is easier.
The agreement between the 2-D work of BDMPS and Melott \& Shandarin's 3-D
work (1993, hereafter MS) is very impressive, suggesting 2-D is a good guide 
to what will happen in 3-D.  The technique can be easily 
applied to two-dimensional galaxy samples, as well as to the microwave 
temperature fluctuation maps.  After testing this method in 2-D we plan to
study more realistic 3-D distributions.

\section{Numerical Simulations}

	The numerical models are the ones used in the work of BDMPS.  They are 
$512^{2}$
density arrays that simulate various epochs of gravitational clustering.
The models are evolved with a particle-mesh code by solving Poisson's
equation and are equivalent to a cross-section of a three-dimensional 
$\Omega=1$
FRW universe with two-dimensional density perturbations.  Thus, they exhibit
a subset of the behavior possible in three dimensions.  The initial conditions
for these models are pure power-law spectra with $n=2$, $n=0$ and $n=-2$, and
are analogous to $n=1$, $n=-1$ and $n=-3$, respectively, in three dimensions.
For more detailed information see BDMPS.  The $n=-2$ model is analogous to 
the CDM model on
small scales, and the $n=0$ model is analogous to the part of the CDM spectrum
reaching non-linearity at $z \sim 0$.  The $n=2$ spectrum is analogous
to that part of the spectrum coming within the horizon at any moment
in most inflationary theories of the generation of perturbations.  
For each type of initial condition we 
evolve density arrays
representing progressive epochs by introducing a non-linearity frequency 
$k_{NL}$, which corresponds to the transition into non-linear evolution.
$k_{NL}$ is defined by:
\[ D^{2}(t) \int_{0}^{k_{NL}}P(k)d^{2}k=1, \]
where $D(t)$ is the growing mode of gravitational instability, 
$D(t) = a(t)$ for the $\Omega=1$ universe, $a(t)$ is the cosmological 
scale factor and, in the absence of
pressure and radiation terms, $a(t)\propto t^{2/3}$.
The non-linearity frequency is chosen to be $k_{NL}=2^{r}k_{f}$ 
where $r=2,3,\ldots,7$
and $k_{f}=2\pi/L$ where $L$ is the size of the simulation box 
(512 grid units).
The Nyquist frequency for these simulations is $k_{N}=256k_{f}$.  
Additionally, we have run four realizations of each model
at each epoch in order to average our results.  We have,
therefore, four realizations for each of six epochs in both the $n=2$ and
$n=0$ simulations and for each of five epochs in the $n=-2$.
We also decided to study a pure Gaussian distribution for the $n=2$ and
$n=0$ models.  For both models we ran ten realizations of a Gaussian
distribution.  Thus we have a total of
eighty-eight files studied, averaged into nineteen sets of results.

\vfill\eject

\section{Smoothing}

	We smooth each density distribution by Fourier convolution with 
the Gaussian window
\[ W(R)=e^{-R^{2}/2R_{S}^{2}}. \]
We specify the smoothing length for each distribution as $R_{S}=R_{1}$ 
where
$R_{1}$ is the
value for which the RMS density fluctuation $\delta\rho/\rho = 1$.  As reported
in BDMPS, $R_{1}=0.8k_{NL}^{-1}$ is extremely stable and is more reliable than
smoothing with the correlation length.  Another advantage of this
choice is that $R_{1}$ in our universe is about $8h^{-1} Mpc$, close to the
mean galaxy separation.
Thus the smoothing will remove most noise due to discreteness effects, and 
may somewhat lessen the difference between the coordinate and redshift spaces.

\section{Percolation}

	Percolation theory has been a tool available to physicists for some
time.  It is used mainly to study phase transitions such as the spontaneous
alignment of spin occurring at the Curie temperature.  Zel'dovich (1982) and
Shandarin (1983) were the first to apply percolation to cosmology, using it
as a tool to study topological properties of non-linear density fields.  In
essence, we use percolation to study phase transitions also.  In our case
however, it is not the system under study that is evolving, we have already
evolved our simulations to the various points in time or epochs we wish to
study.  Rather, we take each of our 'snap shots' of gravitational clustering
and allow the 'threshold density' to change.  We choose an initial threshold
density and increment or decrement this density until percolation in the
over-dense or under-dense phase is reached using a square-lattice percolation
 algorithm.  The
percolation threshold is the point at which the transition is made from
discontinuous clusters to a global cluster which spans the simulation box.
The density at which this threshold is reached is called the percolation
density.  This percolation density is applied to the smoothed density files
such that anything equal to or above percolation density becomes a 1 and 
anything below becomes a 0.  We call this our percolated file.  For the sake
of simplicity, we will call regions above the threshold 'superclusters' and
those below it 'voids.'  We wish to stress that these are not necessarily
identical to the observationally based use of these words, and emphasize 
this by using single quotes.

	In the ideal continuous 2-D world, percolation in the over-dense
phase would mean the absence of percolation in the under-dense phase and
vice-versa.  However, we study percolation on a 2-D square lattice, and 
therefore it may be possible that neither phase percolates.  In order to make
the procedure symmetric, we apply the void search at the percolation threshold
in the over-dense phase and, measuring the 'supercluster' sizes, we use the
percolation threshold in the under-dense phase.  This technique can be used
to study the largest voids when they are isolated, since the over-dense phase
percolates and vice-versa.  One can find a more detailed
discussion of percolation in 2-D systems in Dominik and Shandarin (1992).

\section{Void Search}

	We perform a 'void' search on the percolated file.  The 'void' search
algorithm (KM) first finds the largest square of empty cells in the percolated
file.  Any empty cells along its four sides are then added to this base 'void',
subject to the criterion that the length of the fill-in cannot be less than
two-thirds the base 'void' length.
When the fill-in is complete, the next largest square is found
and cells are added to it.  This process continues for smaller and smaller
squares until all the 'voids' in a particular simulation are found.  We weight
each 'void' size by its predominance in the simulation.  We characterize each
simulation by the diameter of a circle of area equal to the weighted 
average 'void' size.
\[ <D> = {{\Sigma dW_{d}} \over {\Sigma W_{d}}} \]
where the summations are over the range of 'void' sizes, 
$d=2, 3, \ldots , d_{MAX}$ and
\[ W_{d} = {{N_{d} d^{2}} \over {d_{MAX}^{2}}} \]
is the weight of 'voids' of size $d$.  $N_{d}$ is the number of 'voids' of 
size $d$.
	For purposes of comparison we performed a 'supercluster' search by
percolating along under-dense regions so that the 'void' search was 
actually finding the typical sizes of the 'superclusters.'

\section{Results and Discussion}

	We present the results in the form of the percentage area occupied
by 'voids' or 'superclusters' as a function of the diameter in figures 1-3,
for $n=2, 0, -2$ respectively.  Panel (a) in figure 1 and 2 shows
the distribution for the Gaussian fields with the corresponding power spectra.
One can see that in Gaussian fields there is no statistical difference 
between the 'void' and 'supercluster' distributions, which was of course
anticipated.  It is worth stressing that the total area in 'voids' and
'superclusters' is only $\approx 12 \%$ in the $n=2$ model and $\approx 31\%$
in the $n=0$ model.  It reflects the fact that the 'void' search algorithm
can measure only large 'voids' (see figure 4).  One can see also that in the
range $D \sim 0.005$, there are about 2.5 times more 'voids' in the $n=0$ 
model than in the $n=2$ model and there are a few larger 'voids' as well.  
Both features are in agreement with the greater smoothness of the $n=0$
model.

	The non-linear stages shown in panels b through g of figure 1 and 2
show the major difference between the $n=2$ and $n=0$ models:  in the $n=2$
model 'superclusters' occupy more area than 'voids' and the largest structures
are also 'superclusters,' in the $n=0$ model the opposite is true.  The 
degenerate $n=-2$ model demonstrates that 'voids' are always the dominant
structures.

	The $n=2$ model shows that both the 'voids' and 'superclusters' 
roughly scale with the scale of non-linearity $\lambda _{NL}$, and the
$n=-2$ model does not, as can be seen in figure 1 through 3 and especially
in figure 5.  The mean diameter of 'superclusters' is slightly but
significantly greater than that of 'voids' in the $n=2$ model, and both
are about 2 times smaller than $\lambda _{NL}$, shown as a dotted line in 
figure 5.  

	In the $n=0$ model 'voids' are a little larger than 'superclusters'
and the difference with $\lambda _{NL}$ is somewhat less, about 1.8 times.
The models also do not display quite as good 
scaling with $\lambda _{NL}$ as the $n=2$ model.

	As previously mentioned in the section on the void search routine,
we don't necessarily include all of the area below the percolation threshold
in 'voids.'  The 'void' search routine was written to approximate circular 
'voids.'
  This is most obvious in figure 4, where the early epochs show a 
combined 'void'/cluster area of considerably less than 100\%.  In the later 
epochs, the 'voids' become larger and more circular; less 'void' area is lost 
due to the two-thirds requirement during fill-in (see description of void 
search).  The careful observer will also notice that a few of the 
data points are slightly above 100\%, most noticably the last two data
points on the $n=-2$ (dashed) line.  This is simply because we measure 
'voids' at the percolation threshold of the over-dense regions and 
'superclusters' at that of the under-dense regions, and these percolation
thresholds are not identical.

	The advantages of our approach are both theoretical and practical.
The smoothing we use removes discreteness effects and our choice of smoothing
length is insensitive to statistical fluctuations.  The use of percolation 
objectively defines a connected structure, and its combination with an
objective measure of void size remedies its major shortcoming as a 
large-scale structure statistic, i.e., it does not give rise to a physical
lengthscale associated with percolating structures.  As shown in earlier
work, the results of percolation studies can be connected with the initial
conditions that generated the structures.

\acknowledgments

	We are grateful for financial support from the University of Kansas
GRF fund, NASA Grant NAGW-2923, and NSF Grants AST-9021414 and OSR-9255223.
PMH is especially grateful for support from the NSF Research Expenses for
Undergraduates program.
	Our computer simulations were done on a grant of Cray-2 time at the
National Center for Supercomputing Applications.
	We wish to thank Guinevere Kauffmann for use of and advice on her
void search software.

\vfill\eject

Figure Captions:

Figure 1a)  Percentage area occupied by 'voids' (solid lines) and 'superclusters'
		(dashed lines) of size L (units of box size) for a purely 
		Gaussian density distribution with $k_{c}=256k_{f}$ in the 
		$n=2$ model.

Figure 1b)  Percentage area occupied by 'voids' (solid lines) and 'superclusters'
		(dashed lines) of size L (units of box size) for the epoch
		$k_{NL}=128$ in the $n=2$ model.

Figure 1c to 1h)  Same as Fig. 1b but $k_{NL}$ as specified on plot.

Figure 2)  Same as Fig. 1 for $n=0$ model

Figure 3a)  Same as Fig. 1 for $n=-2$ model except that only stages
	    $k_{NL}=64, 32, 16, 8, and 4$ are shown.

Figure 4)  Sum of areas occupied by both 'voids' and 'superclusters' versus
		$\lambda_{NL}$ for $n=2$ (solid line), $n=0$ (dotted line)
		and $n=-2$ (dashed line) models.

Figure 5a)  Evolution of mean 'void' size (solid line) and 'supercluster' size 
		(dashed line) versus $\lambda_{NL}$ in the
		$n=2$ model.  $d=\lambda_{NL}$ (dotted line) is included
		for comparison.

Figure 5b)  Same as Fig. 5a but in the $n=0$ model.

Figure 5c)  Same as Fig. 5a but in the $n=-2$ model.

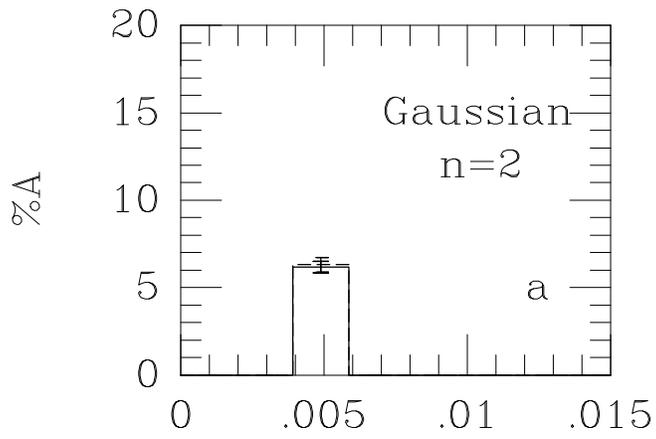
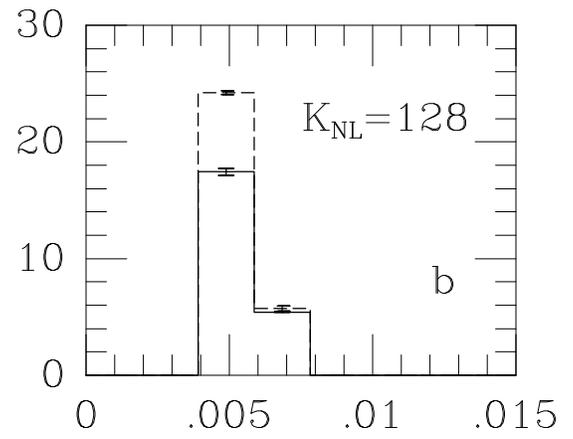
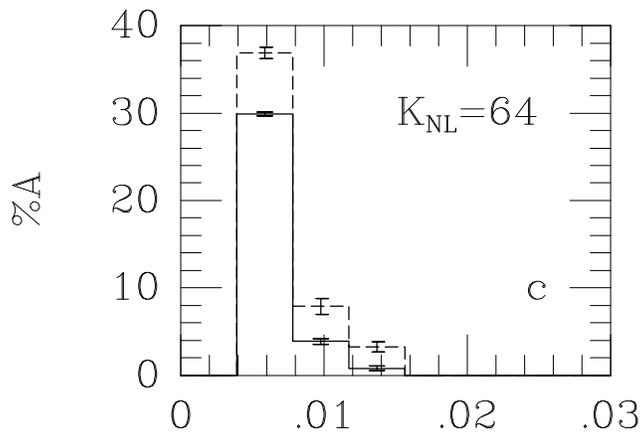
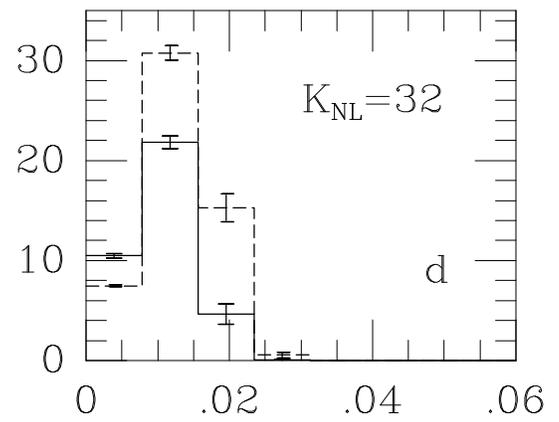
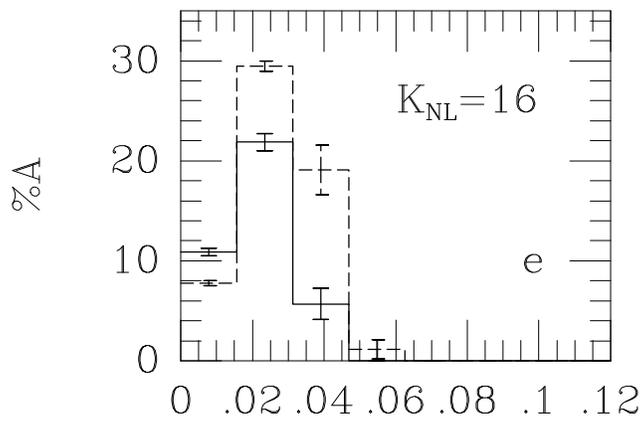
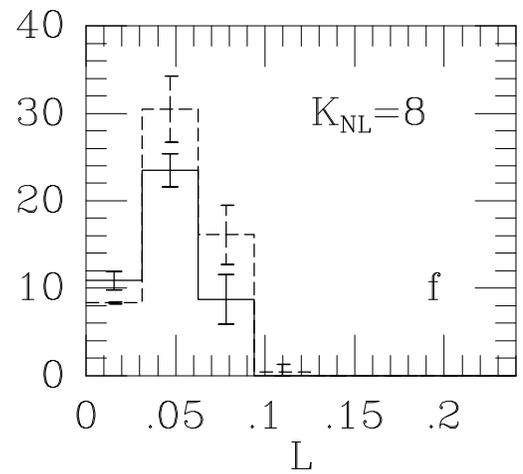
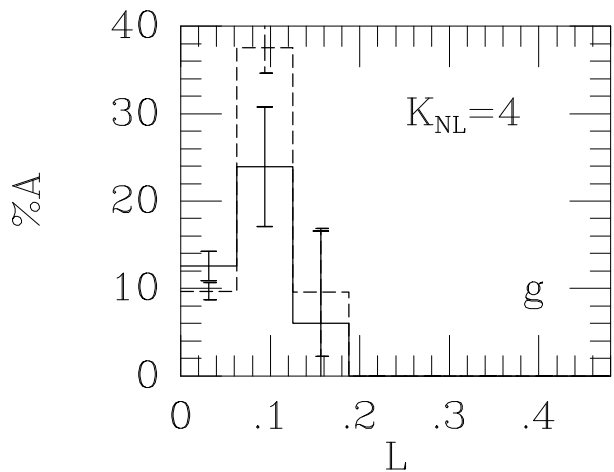
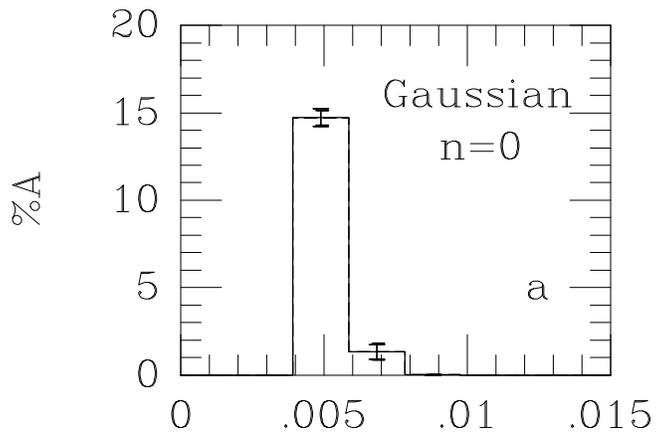
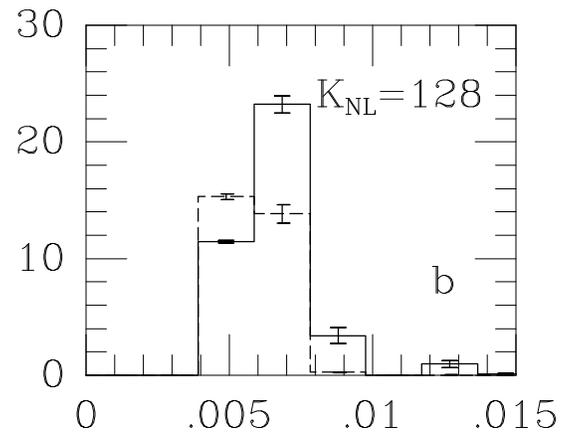
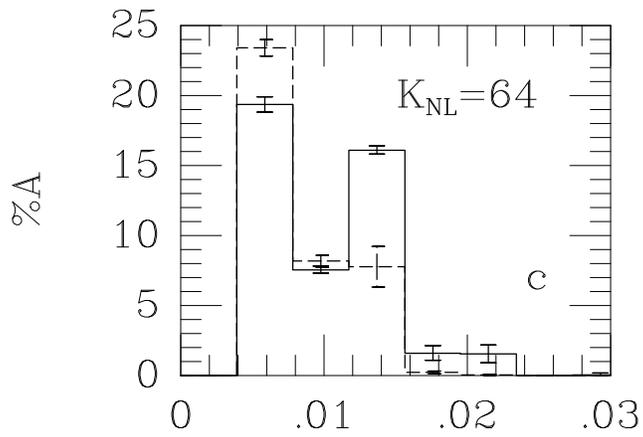
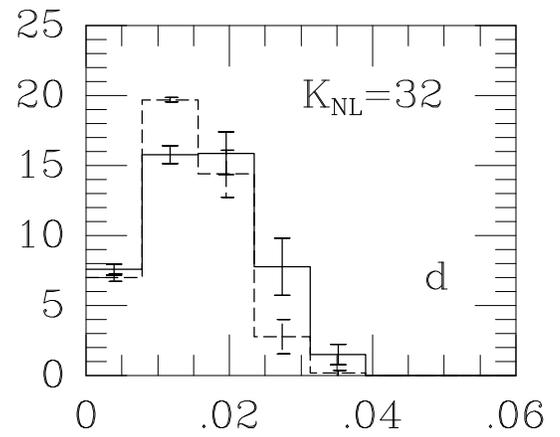
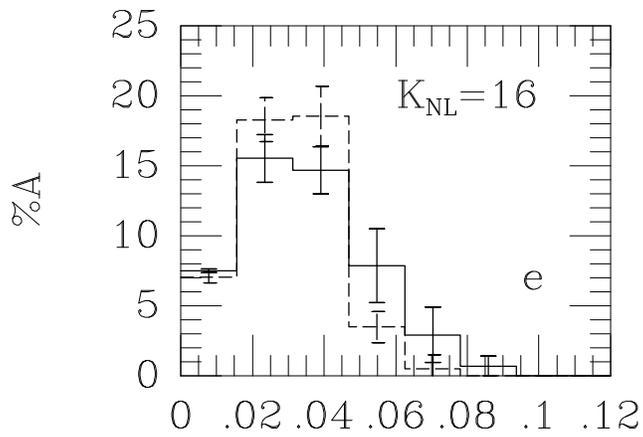
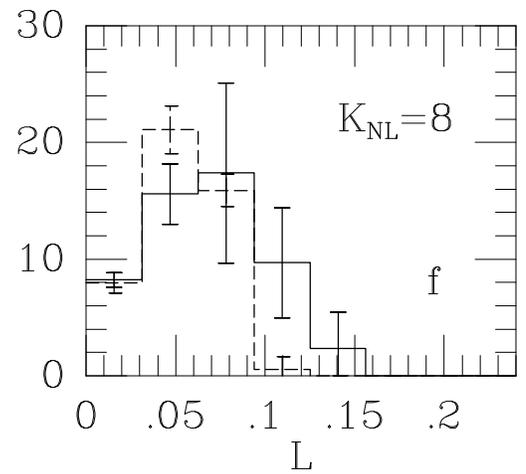
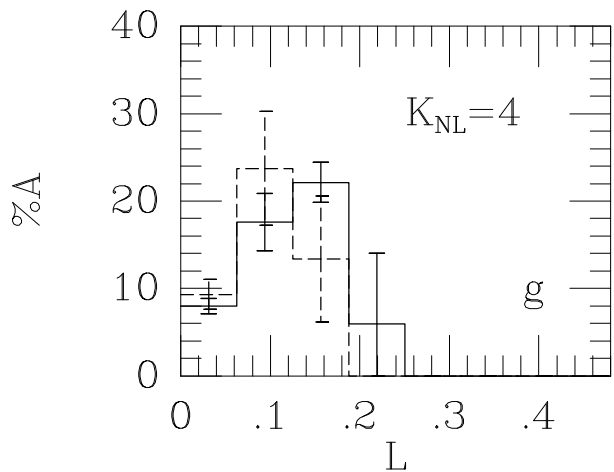
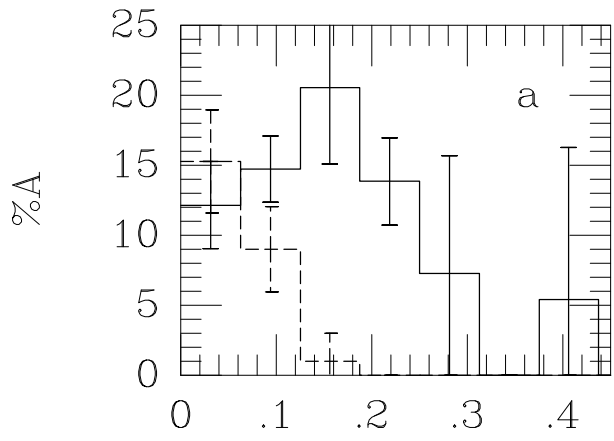
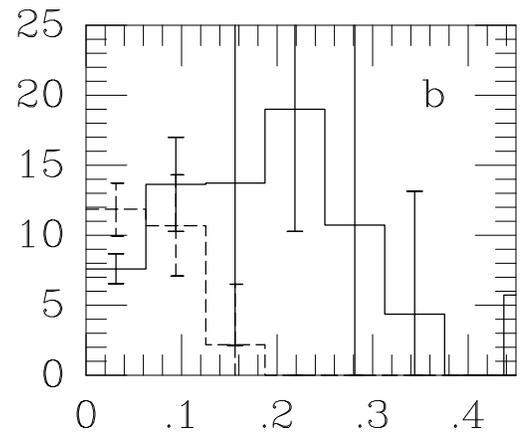
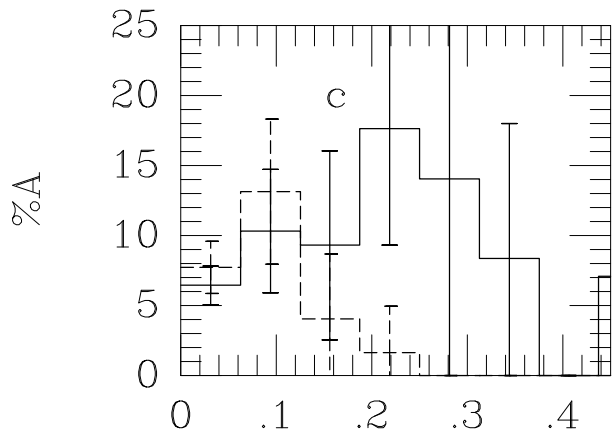
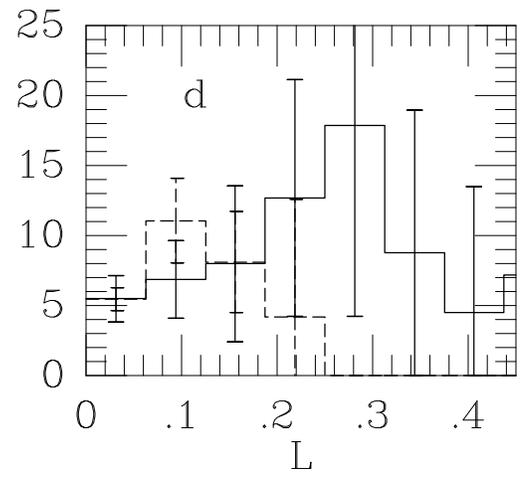
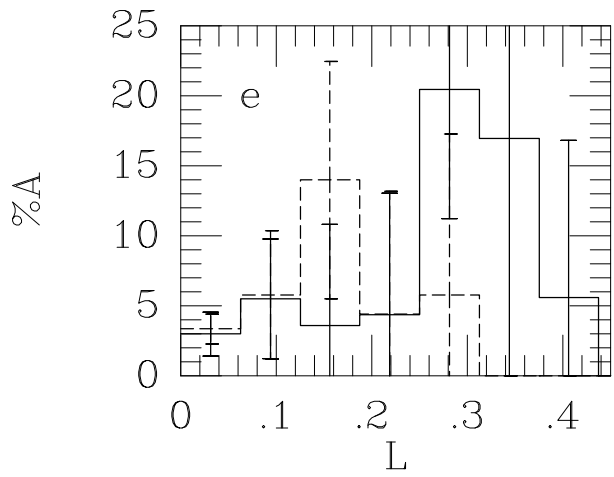
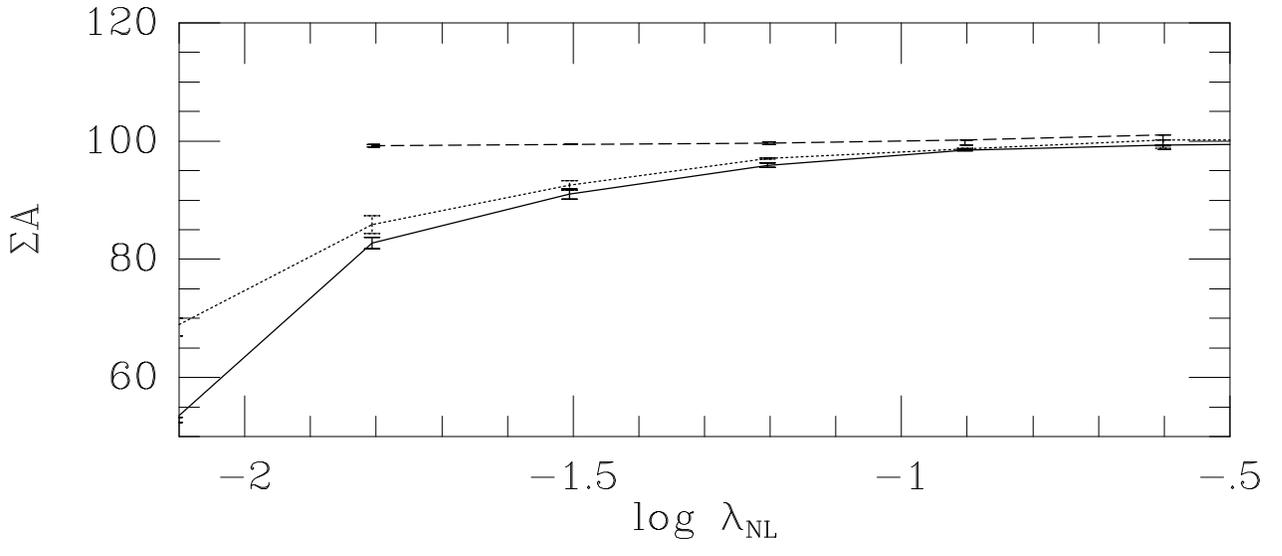
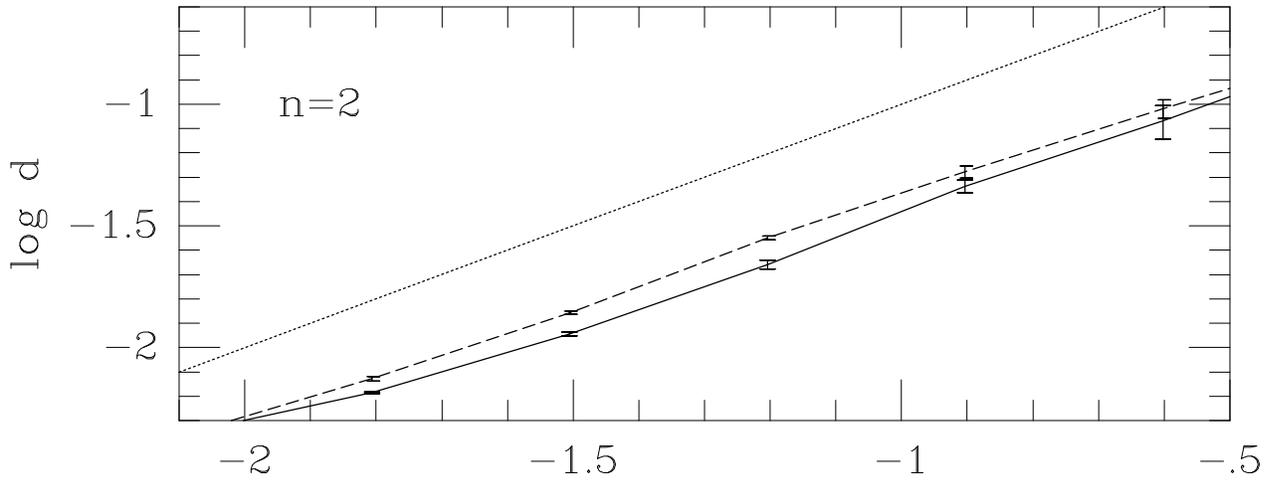
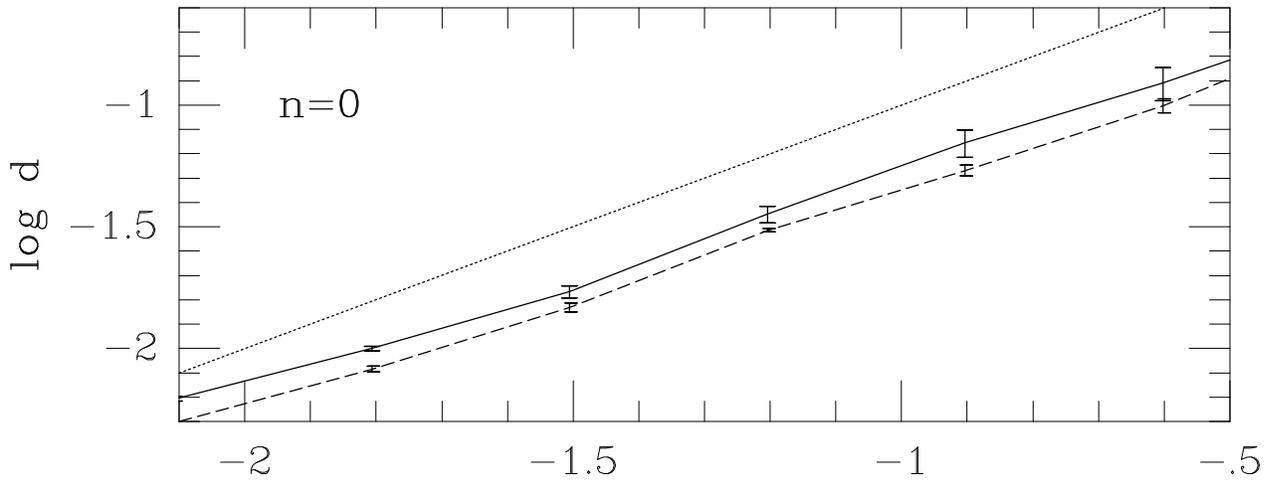
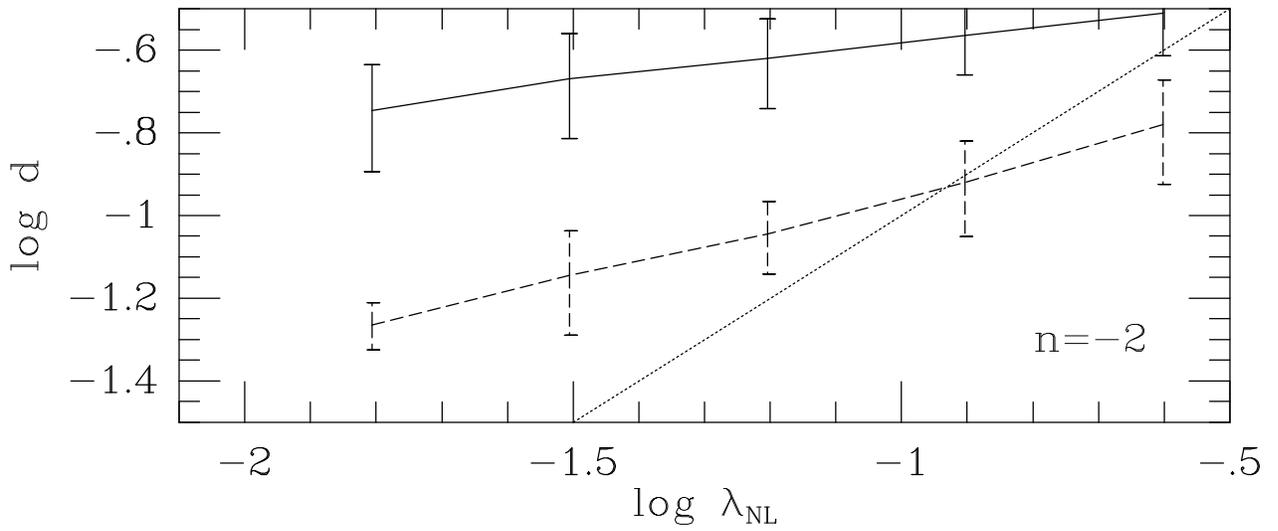
\end{document}